\documentclass[runningheads]{llncs}
\usepackage{graphicx}
\usepackage{tikz}
\usepackage{comment} 
\usepackage{amsmath,amssymb} 
\usepackage{color}
\usepackage{graphicx}
\usepackage{mwe}
\usepackage{xcolor,colortbl}
\usepackage{tabularx}
\usepackage{multirow}
\usepackage{relsize}
\usepackage{multirow}
\usepackage{adjustbox}
\usepackage{acro}
\usepackage{tikz}
\usepackage{comment} 
\usepackage{amsmath,amssymb} 
\usepackage{color}
\usepackage{graphicx}
\usepackage{comment}
\usepackage{amsmath,amssymb} 
\usepackage{color}
\usepackage{float}
\usepackage{epsfig}
\usepackage{graphicx}
\usepackage{amsmath}
\usepackage{amssymb}
\usepackage{mwe}
\usepackage{xcolor,colortbl}
\usepackage{tabularx}
\usepackage{multirow}
\usepackage{relsize}
\usepackage{pifont}
\usepackage{multirow}
\usepackage{adjustbox}
\usepackage{acro}

\pdfoutput=1

\DeclareAcronym{CT}{
short=CT,
long=computed tomography,
long-plural-form = computed tomographies,
 foreign-plural={}
}

\DeclareAcronym{ASSD}{
short=ASSD,
long=average symmetric surface distance
}

\DeclareAcronym{ICC}{
short=ICC,
long=intrahepatic cholangiocellular carcinoma
}

\DeclareAcronym{HCC}{
short=HCC,
long=hepatocellular carcinoma
}

\DeclareAcronym{HeMIS}{
short=HeMIS,
long=hetero-modality image segmentation
}

\DeclareAcronym{PACS}{
short=PACS,
long=picture archiving and communication system,
 foreign-plural = {}
}

\DeclareAcronym{TACE}{
short=TACE,
long=transarterial chemoembolization
}
\DeclareAcronym{CGMH}{
short=CGMH,
long=Chang Gung Memorial Hospital
}
\DeclareAcronym{ADA}{
short=ADA,
long=adversarial domain adaptation
}

\DeclareAcronym{PET}{
short=PET,
long=positron emission tomography
}
\DeclareAcronym{FCN}{
short=FCN,
long=fully convolutional network,
 foreign-plural = {}
}

\DeclareAcronym{A}{
short=A,
long=arterial
}
\DeclareAcronym{SGD}{
short=SGD,
long=stochastic gradient descent
}
\DeclareAcronym{V}{
short=V,
long=venous
}

\DeclareAcronym{JSD}{
short=JSD,
long=Jensen-Shannon divergence
}

\DeclareAcronym{D}{
short=D,
long=delay
}

\DeclareAcronym{NC}{
short=NC,
long=non-contrast
}

\DeclareAcronym{CHASe}{
short=CHASe,
long=co-heterogenous and adaptive segmentation
}

\DeclareAcronym{DSC}{
short=DSC,
long=Dice-S\o{}rensen coefficient,
 foreign-plural={}
}

\DeclareAcronym{PHNN}{
short=PHNN,
long=progressive holistically nested network
}
\DeclareAcronym{ASPP}{
short=ASPP,
long=atrous spatial pyramid pooling
}

\newcommand{\NC}{\mathit{NC}}
\newcommand{\cmark}{\ding{51}}
\newcommand{\xmark}{\ding{54}}
\newcommand*\rot{\rotatebox{90}}
\newcommand{\Sec}{Sec.}
\newcommand{\Tab}{Table}
\newcommand{\Fig}{Figure}
\newcommand{\eg}{\textit{e.g.},}
\newcommand{\ie}{\textit{i.e.},}
\newcommand{\etal}{\textit{et al.}}

\begin{document}
\pagestyle{headings}
\mainmatter
\def\ECCVSubNumber{4353}  

\title{Co-Heterogeneous and Adaptive Segmentation from Multi-Source and Multi-Phase CT Imaging Data: A Study on Pathological Liver and Lesion Segmentation} 

\titlerunning{CHASe: Co-Heterogeneous and Adaptive Segmentation}
%
\author{Ashwin Raju\inst{1,2} \and
Chi-Tung Cheng\inst{3} \and
Yuankai Huo\inst{1} \and
Jinzheng Cai\inst{1} \and
Junzhou Huang\inst{2} \and
Jing Xiao\inst{4} \and
Le Lu\inst{1} \and
ChienHung Liao\inst{3}\and
Adam P. Harrison\inst{1}}

\authorrunning{A. Raju et al.}
%
\institute{PAII Inc., Bethesda MD, USA \and
The University of Texas at Arlington, Arlington TX, USA \and
Chang Gung Memorial Hospital, Linkou, Taiwan, ROC\and
PingAn Technology, Shenzhen, China
}
\maketitle

\begin{abstract}
Within medical imaging, organ/pathology segmentation models trained on current publicly available and fully-annotated datasets usually do not well-represent the heterogeneous modalities, phases, pathologies, and clinical scenarios encountered in real environments. On the other hand, there are tremendous amounts of unlabelled patient imaging scans stored by many modern clinical centers. In this work, we present a novel segmentation strategy, \ac{CHASe}, which only requires a small labeled cohort of \textit{single} phase data to adapt to any unlabeled cohort of heterogenous \textit{multi-phase} data with possibly new clinical scenarios and pathologies. To do this, we develop a versatile framework that fuses appearance-based semi-supervision, mask-based \acl{ADA}, and pseudo-labeling. We also introduce co-heterogeneous training, which is a novel integration of co-training and hetero-modality learning. We evaluate \ac{CHASe} using a clinically comprehensive and challenging dataset of multi-phase \ac{CT} imaging studies ($1147$ patients and $4577$ 3D volumes), with a test set of $100$ patients. Compared to previous state-of-the-art baselines, \ac{CHASe} can further improve pathological liver mask \aclp{DSC} by ranges of $4.2\%$ to $9.4\%$, depending on the phase combinations, \eg{} from $84.6\%$ to $94.0\%$ on non-contrast CTs. 
\keywords{multi-phase segmentation, semi-supervised learning,  co-training, domain adaptation, liver and lesion segmentation}
\end{abstract}
\acresetall

\section{Introduction}

Delineating anatomical structures is an important task within medical imaging, \eg{} to generate biomarkers, quantify or track disease progression, or to plan radiation therapy. Manual delineation is prohibitively expensive, which has led to a considerable body of work on automatic segmentation. However, a perennial problem is that models trained on available image/mask pairs, \eg{} publicly available data, do not always reflect clinical conditions upon deployment, \eg{} the present pathologies, patient characteristics, scanners, and imaging protocols. This leads to potentially drastic performance gaps. When multi-modal or multi-phase imagery is present these challenges are further compounded, as datasets may differ in their composition of available modalities or consist of heterogeneous combinations of modalities. The challenges then are in both managing new patient/disease variations and in handling heterogeneous multi-modal data. Ideally segmentation models can be deployed without first annotating large swathes of additional data matching deployment scenarios.  This is our goal, where we introduce \ac{CHASe}, which can adapt models trained on single-modal and public data to produce state-of-the-art results on \textit{multi-phase and multi-source} clinical data \textit{with no extra annotation cost}.  
\begin{figure}[t]
     \centering
    
      \includegraphics[width=.8\linewidth]{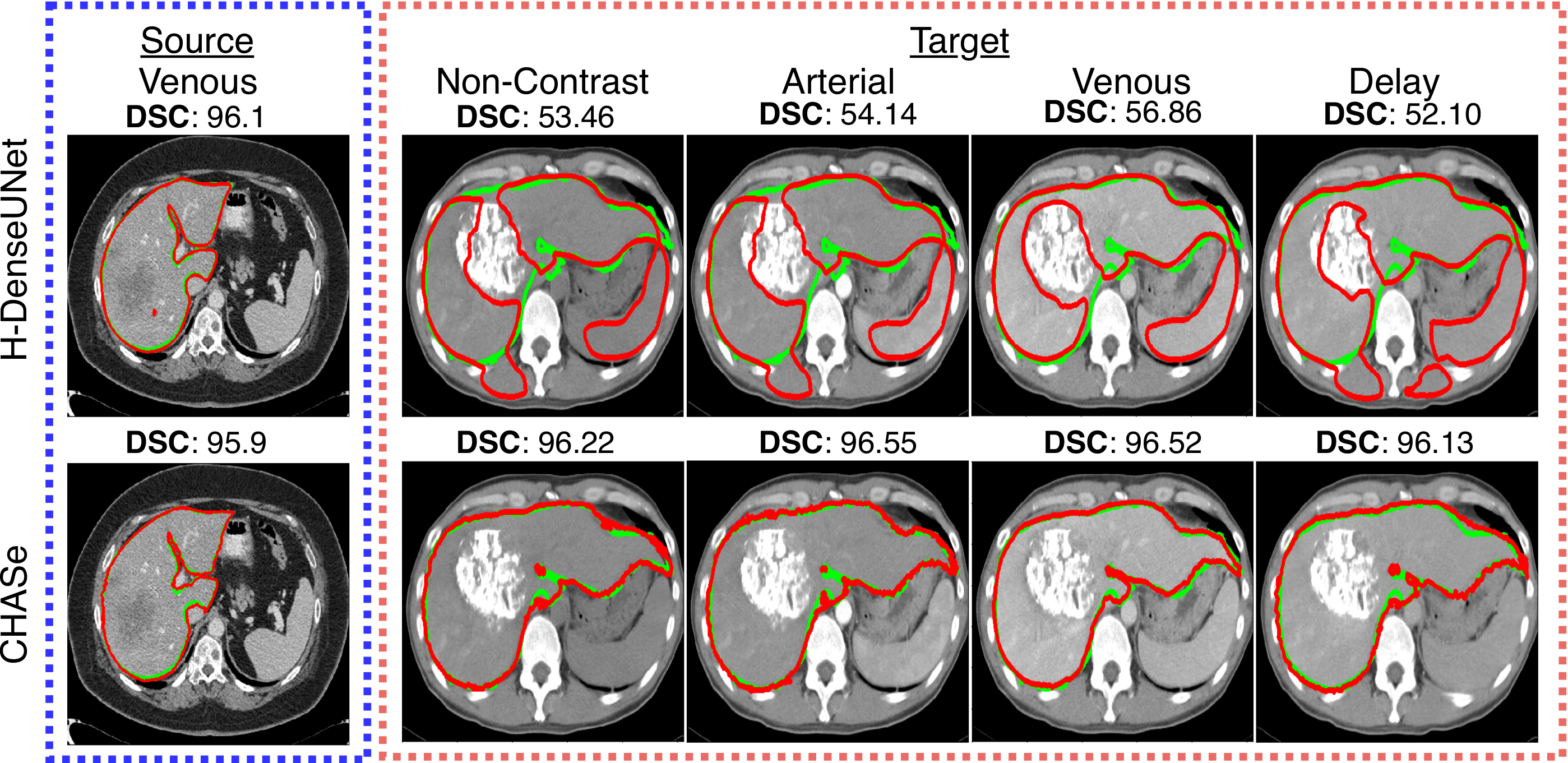}

    \caption{Ground truth and predictions are rendered in green and red, respectively. Despite performing excellently on labeled source data, state-of-the-art fully-supervised models~\cite{li2018h} can struggle on cohorts of multi-phase data with  novel conditions, \eg{} the patient shown here with splenomegaly and a \acs{TACE}-treated tumor. \ac{CHASe} can adapt such models to perform on new data.}
    \label{fig:intro}
\end{figure}

Our motivation stems from challenges in handling the wildly variable data found in \acp{PACS}, which closely follows deployment scenarios. In this study, we focus on segmenting pathological livers and lesions from dynamic \ac{CT}. Liver disease and  cancer are major morbidities, driving efforts toward better detection and characterization methods. The dynamic contrast \ac{CT} protocol images a patient under multiple time-points after a contrast agent is injected, which is critical for characterizing liver lesions~\cite{Maria_2004}.  Because accurate segmentation produces important volumetric biomarkers~\cite{gotra_liver_2017,bilic_liver_2019}, there is a rich body of work on automatic segmentation~\cite{milletari2016v,isensee2018nnu,8045995,li2018h,zhanglight,8035318,vorontsov2014metastatic,conze2017scale,roth2019liver}, particularly for \ac{CT}. Despite this, all publicly available data~\cite{bilic_liver_2019,4781564,gibson_eli_2018_1169361,soler20103d,chaos} is limited to venous-phase (single-channel) \acp{CT}. Moreover, when lesions are present, they are typically limited to \ac{HCC} or metastasized tumors, lacking representation of \ac{ICC} or the large bevy of benign lesion types. Additionally, public data may not represent other important scenarios, \eg the \acf{TACE} of lesions or splenomegaly, which produce highly distinct imaging patterns. As \Fig~\ref{fig:intro} illustrates, even impressive leading entries~\cite{li2018h} within the public LiTS challenge~\cite{bilic_liver_2019}, can struggle on clinical \ac{PACS} data, particularly when applied to non-venous contrast phases.

To meet this challenge, we integrate together powerful, but complementary, strategies: hetero-modality learning, appearance-based consistency constraints, mask-based \ac{ADA}, and pseudo-labeling. The result is a semi-supervised model trained on smaller-scale supervised public \textit{venous-phase} data~\cite{bilic_liver_2019,4781564,soler20103d,gibson_eli_2018_1169361,chaos} and large-scale unsupervised \textit{multi-phase} data. Crucially, we articulate non-obvious innovations that avoid serious problems from a naive integration. A key component is co-training~\cite{blum1998combining}, but unlike recent deep approaches~\cite{xia20183d,zhou2019semi}, we do not need artificial views, instead treating each phase as a view. We show how co-training can be adopted with a minimal increase of parameters. Second, since \ac{CT} studies from clinical datasets may exhibit any combination of phases, ideally liver segmentation should also be able to accept whatever combination is available, with performance topping out as more phases are available. To accomplish this, we fuse hetero-modality learning~\cite{havaei2016hemis} together with co-training, calling this \textit{co-heterogeneous training}. Apart from creating a natural hetero-phase model, this has the added advantage of combinatorially increasing the number of views for co-training from $4$ to $15$, boosting even single-phase performance. To complement these appearance-based semi-supervision strategies, we also apply pixel-wise \ac{ADA}~\cite{tsai2018learning}, guiding the network to predict masks that follow a proper shape distribution. Importantly, we show how \ac{ADA} can be applied to co-heterogeneous training with no extra computational cost over adapting a single phase. Finally, we address edge cases using a principled pseudo-labelling technique specific to pathological organ segmentation. These innovations combine to produce a powerful approach we call \ac{CHASe}. 

We apply \ac{CHASe} to a challenging unlabelled dataset of $1147$ dynamic-contrast \ac{CT} studies of patients, all with liver lesions. The dataset, extracted directly from a hospital \ac{PACS}, exhibits many features not seen in public single-phase data and consists of a heterogeneous mixture of \ac{NC}, \ac{A}, \ac{V}, and \ac{D} phase \acp{CT}. With a test set of $100$ studies, \textit{this is the largest, and arguably the most challenging, evaluation of multi-phase pathological liver segmentation to date}. Compared to strong fully-supervised baselines~\cite{li2018h,harrison2017progressive} only trained on public data, \ac{CHASe} can dramatically boost segmentation performance on non-venous phases, \eg{} moving the mean \ac{DSC} from $84.5$ to $94.0$ on \ac{NC} \acp{CT}. Importantly, performance is also boosted on \ac{V} phases, \ie{} from $90.7$ mean \ac{DSC} to $94.9$. Inputting all available phases to \ac{CHASe} maximizes performance, matching desired behavior. Importantly, \ac{CHASe} also significantly improves robustness, operating with much greater reliability and without deleterious outliers. Since \ac{CHASe} is general-purpose, it can be applied to other datasets or even other organs with ease.  

\section{Related Work} \label{related_works}

{\bf Liver and liver lesion segmentation.}
In the past decade, several works addressed liver and lesion segmentation using traditional texture and statistical features~\cite{8035318,vorontsov2014metastatic,conze2017scale}. With the advent of deep learning, \acp{FCN} have quickly become dominant. These include 2D~\cite{roth2019liver,ben2016fully}, 2.5D~\cite{han2017automatic,wang_pairwise_2019}, 3D~\cite{milletari2016v,isensee2018nnu,8045995,yang2017automatic}, and hybrid~\cite{li2018h,zhanglight} \ac{FCN}-like architectures. Some reported results show that 3D models can improve over 2D ones, but these improvements are sometimes marginal~\cite{isensee2018nnu}. 

Like related works, we also use \acp{FCN}. However, all prior works are trained and evaluated on venous-phase \acp{CT} in a fully-supervised manner. In contrast, we aim to robustly segment a large cohort of multi-phase \ac{CT} \ac{PACS} data in a semi-supervised manner. As such, our work is orthogonal to much of the state-of-the-art, and can, in principle, incorporate any future fully-supervised solution as a starting point. 

{\bf Semi-Supervised Learning.}
Annotating medical volumetric data is time consuming, spurring research on semi-supervised solutions~\cite{tajbakhsh2019embracing}. In \textit{co-training}, predictions of different ``views'' of the same unlabelled sample are enforced to be consistent~\cite{blum1998combining}. Recent works integrate co-training to deep-learning~\cite{qiao2018deep,xia20183d,zhou2019semi}. While \ac{CHASe} is related to these works, it uses different contrast phases as views and therefore has no need for artificial view creation~\cite{qiao2018deep,xia20183d,zhou2019semi}. More importantly, \ac{CHASe} effects a stronger appearance-based semi-supervision  by fusing co-training with hetero-modality learning (co-heterogenous learning). In addition, \ac{CHASe} complements this appearance-based strategy via prediction-based \ac{ADA}, resulting in significantly increased performance.

\textit{\Acf{ADA}} for semantic segmentation has also received recent attention in medical imaging~\cite{tajbakhsh2019embracing}. The main idea is to align the distribution of predictions or features between the source and target domains, with many successful approaches~\cite{tsai2018learning,vu2019advent,chen2019crdoco,chang2019all}. Like Tsai \etal{}, we align the distribution of \textit{mask shapes} of source and target predictions~\cite{tsai2018learning}. Unlike Tsai \etal{}, we use \ac{ADA} in conjunction with appearance-based semi-supervision. In doing so, we show how \ac{ADA} can effectively adapt $15$ different hetero-phase predictions at the same computational cost as a single-view variant. Moreover, we demonstrate the need for complementary semi-supervised strategies in order to create a robust and practical medical segmentation system. 

In \textit{self-learning},  a network is first trained with labelled data. The trained model then produces pseudo labels for the unlabelled data, which is then added to the labelled dataset via some scheme~\cite{Lee_2013}. This approach has seen success within medical imaging~\cite{zhang2018self,bai2017semi,min2019two}, but it is important to guard against ``confirmation bias''~\cite{li2019decoupled}, which can compound initial misclassifications~\cite{tajbakhsh2019embracing}. While we thus avoid this approach, we do show later that co-training can also be cast as self-learning with consensus-based pseudo-labels. Finally, like more standard self-learning, we also generate pseudo-labels to finetune our model. But, these are designed to deduce and correct likely mistakes, so they do not follow the regular ``confirmation''-based self-learning framework. 

{\bf Hetero-Modality Learning.}
In medical image acquisition, multiple phases or modalities are common, \eg{} dynamic \ac{CT}. It is also common to encounter hetero-modality data, \ie{} data with possible missing modalities~\cite{havaei2016hemis}. Ideally a segmentation model can use whatever phases/modalities are present in the study, with performance improving the more phases are available. \ac{CHASe} uses hetero-modal fusion for fully-supervised \acp{FCN}~\cite{havaei2016hemis}; however it fuses it with co-training, thereby using multi-modal learning to perform appearance-based learning from \textit{unlabelled} data. Additionally, this \emph{co-heterogeneous training} combinatorially increases the number of views for co-training, significantly boosting even single-phase performance by augmenting the training data. To the best of our knowledge, we are the first to propose co-heterogenous training.


\section{Method}

Although \ac{CHASe} is not specific to any organ, we will assume the liver for this work. We assume we are given a curated and labelled dataset of \acp{CT} and masks, \eg{} from public data sources. We  denote this dataset $\mathcal{D}_{\ell} = \{\mathcal{X}_{i}, Y_{i}\}_{i=1}^{N_{\ell}}$, with $\mathcal{X}_{i}$ denoting the set of available phases and $Y_{i}(k) \in \{0,1,2\}$ indicating background, liver, and lesion for all pixel/voxel indices $k$. Here, without loss of generality, we assume the \acp{CT} are all \ac{V}-phase, \ie{} $\mathcal{X}_{i}=V_{i} \,\,\forall \mathcal{X}_{i} \in  \mathcal{D}_{\ell}$. We also assume we are given a large cohort of unlabelled multi-phase \acp{CT} from a challenging and uncurated clinical source, \eg{} a \ac{PACS}. We denote this dataset $\mathcal{D}_{u} = \{\mathcal{X}_{i}\}_{i=1}^{N_{u}}$, where $\mathcal{X}_{i}=\{\NC_{i}, A_{i}, V_{i}, D_{i}\}$ for instances with all contrast phases. When appropriate, we drop the $i$ for simplicity. Our goal is to learn a segmentation model which can accept any combination of phases from the target domain and robustly delineate liver or liver lesions, despite possible differences in morbidities, patient characteristics, and contrast phases between the two datasets.

\begin{figure*}[t]
\begin{center}
        \includegraphics[width=.95\linewidth]{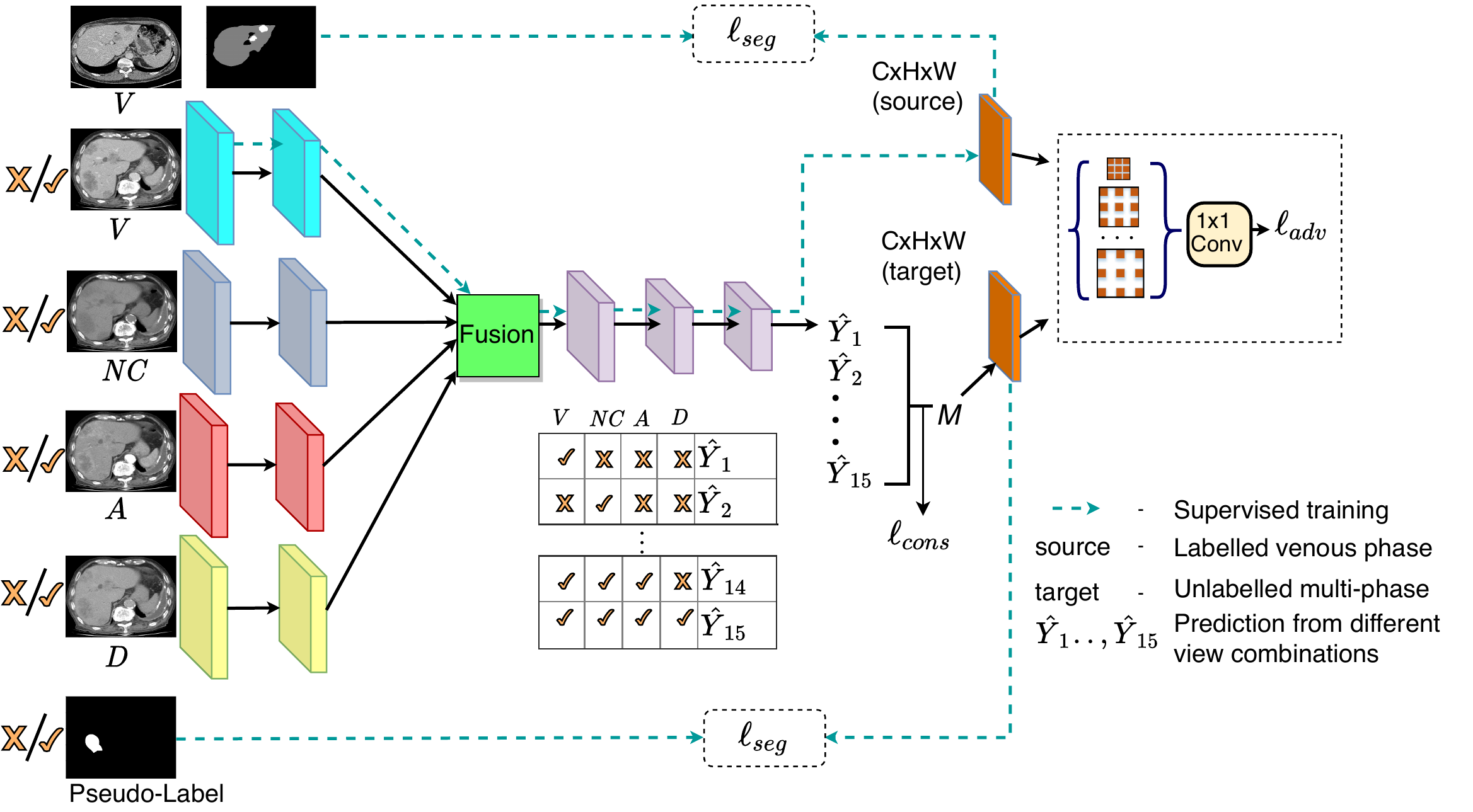}
\end{center}
   \caption{\small \textbf{Overview of \ac{CHASe}.} As shown at the top, labelled data, \ac{V}-phase in our experiments, are trained using standard segmentation losses. Multi-phase unlabelled data are inputted into an efficient co-heterogenous pipeline, that minimizes divergence between mask predictions across all $15$ phase combinations. \acs{ADA} and specialized pseudo-labelling are also applied. Not shown are the deeply-supervised intermediate outputs of the \acs{PHNN} backbone. }
\label{fig:system}
\end{figure*}


\Fig~\ref{fig:system} \ac{CHASe}, which integrates supervised learning, co-heterogenous training, \ac{ADA}, and specialized pseudo-labelling. We first start by training a standard fully-supervised segmentation model using the labelled data. Then under \ac{CHASe}, we finetune the model using consistency and \ac{ADA} losses: 
\begin{align}
   \mathcal{L} = \mathcal{L}_{seg} + \mathcal{L}_{cons} + \lambda_{adv}\mathcal{L}_{adv} \mathrm{,} \label{eqn:total}
\end{align}
where $\mathcal{L}$, $\mathcal{L}_{seg}$, $\mathcal{L}_{cons}$, $\mathcal{L}_{adv}$ are the overall, supervised, co-heterogenous, and \ac{ADA} losses, respectively. For adversarial optimization, a discriminator loss, $\mathcal{L}_{d}$, is also deployed in competition with  \eqref{eqn:total}. We elaborate on each of the above losses below. Throughout, we minimize hyper-parameters by using standard architectures and little loss-balancing. 



\subsection{Backbone}
\label{sec:backbone}
\ac{CHASe} relies on an \ac{FCN} backbone, $f(.)$, for its functionality and can accommodate any leading choice, including encoder/decoder backbones~\cite{RFB15a,milletari2016v}. Instead, we adopt the deeply-supervised \acf{PHNN} framework~\cite{harrison2017progressive}, which has demonstrated leading segmentation performance for many anatomical structures~\cite{harrison2017progressive,roth2018spatial,Cai_2018,Jin_2019a,Jin_2019b}, sometimes even outperforming U-Net~\cite{Jin_2019a,Jin_2019b}. Importantly,  \ac{PHNN} has roughly half the parameters and activation maps of an equivalent encoder/decoder. Since we aim to include additional components for semi-supervised learning, this lightweightedness is important.

In brief, \ac{PHNN} relies on deep supervision in lieu of a decoder and assumes the \ac{FCN} can be broken into stages based on pooling layers. Here, we assume there are five \ac{FCN} stages, which matches popular \ac{FCN} configurations. \ac{PHNN} produces a sequence of logits, $a^{(m)}$, using $1\times1$ convolutions and upsamplings operating on the terminal backbone activations of each stage.  Sharing similarities to residual connections~\cite{He2015DeepRL}, predictions are generated for each stage using a progressive scheme that adds to the previous stage's activations:
\begin{align}
    \hat{Y}^{(1)} &= \sigma(a^{(1)}) \mathrm{,} \\
    \hat{Y}^{(m)} &= \sigma(a^{(m)}+a^{(m-1)}) \,\,\forall m>1 \mathrm{,}
\end{align}
where $\sigma(.)$ denotes the softmax and $\hat{Y}^{(.)}$ represents the predictions, with the final stage's predictions acting as the actual segmentation output, $\hat{Y}$. Being deeply supervised, \ac{PHNN} optimizes a loss at each stage, with higher weights for later stages:
\begin{align}
    \ell_{seg}(f(V), Y) = \sum_{j=1}^{5}\dfrac{m}{5}\,\ell_{ce}\left(\hat{Y}^{m}, Y\right) \mathrm{,}
\end{align}
where we use pixel-wise cross-entropy loss (weighted by prevalence), $\ell_{ce}(.,.)$. Prior to any semi-supervised learning, this backbone is  pre-trained using $\mathcal{D}_{\ell}$: 
\begin{align}
    \mathcal{L}_{seg} = \dfrac{1}{N_{\ell}}\sum_{V,Y\in\mathcal{D}_{\ell}} \ell_{seg}\left(f(V), Y\right) \label{eqn:seg} \mathrm{.}
\end{align}
Our supplementary visually depicts the \ac{PHNN} structure.



\subsection{Co-Training}
\label{sec:co-train}
With a pretrained fully-supervised backbone, the task is now to leverage the unlabelled cohort of dynamic \ac{CT} data, $\mathcal{D}_{u}$. We employ the ubiquitous strategy of enforcing consistency. Because dynamic \ac{CT} consists of the four \ac{NC}, \ac{A}, \ac{V}, and \ac{D} phases, each of which is matched to same mask, $Y$, they can be regarded as different views of the same data. This provides for a natural co-training setup~\cite{blum1998combining} to penalize inconsistencies of the mask predictions across different phases. 

To do this, we must create predictions for each phase. As \Fig~\ref{fig:system} illustrates, we accomplish this by using phase-specific \ac{FCN} stages, \ie{} the first two low-level stages, and then use a shared set of weights for the later semantic stages. Because convolutional weights are more numerous at later stages, this allows for an efficient multi-phase setup. All layer weights are initialized using the fully-supervised \ac{V}-phase weights from \Sec~\ref{sec:backbone}, including the phase-specific layers. Note that activations across phases remain distinct. Despite the distinct activations, for convenience we abuse notation and use $\mathcal{\hat{Y}}=f(\mathcal{X})$ to denote the generation of all phase predictions for one data instance. When all four phases are available in $\mathcal{X}$, then $\mathcal{\hat{Y}}$ corresponds to $\{\hat{Y}^{\NC}$, $\hat{Y}^{A}$, $\hat{Y}^{V}$, $\hat{Y}^{D}\}$.

Like Qiao \etal{}~\cite{qiao2018deep} we use the \ac{JSD}~\cite{lin1991divergence} to penalize inconsistencies. However, because it will be useful later we devise the \ac{JSD} by first deriving a consensus prediction:
\begin{align}
    M = \dfrac{1}{|\mathcal{\hat{Y}}|}\sum_{\hat{Y}\in\mathcal{\hat{Y}}}\hat{Y} \mathrm{.} \label{eqn:M}
\end{align}
The \ac{JSD} is then the divergence between the consensus and each phase prediction:
\begin{align}
    \ell_{cons}(f(\mathcal{X})) &= \dfrac{1}{|\mathcal{\hat{Y}}|}\sum_{\hat{Y}\in \mathcal{\hat{Y}}}  \sum_{k \in \Omega}\mathit{KL}(\hat{Y}(k) \parallel M(k)) \mathrm{,} \label{eqn:co_instance_1} \\
    \mathcal{L}_{cons} &= \dfrac{1}{N_{u}}\sum_{\mathcal{X}\in\mathcal{D}^{u}} \ell_{cons}(f(\mathcal{X})) \mathrm{,} \label{eqn:co_1}
\end{align}
where $\Omega$ denotes the spatial domain and $\mathit{KL}(.\parallel.)$ is the Kullback-Leibler divergence across label classes. \eqref{eqn:co_1} thus casts co-training as a form of self-learning, where pseudo-labels correspond to the consensus prediction in \eqref{eqn:M}. For the deeply-supervised \ac{PHNN}, we only calculate the \ac{JSD} across the final prediction.




\subsection{Co-Heterogeneous Training}
\label{sec:co-hemis}

While the loss in \eqref{eqn:co_1} effectively incorporates multiple phases of the unlabelled data, it is not completely satisfactory. Namely, each phase must still be inputted separately into the network, and there is no guarantee of a consistent output. Despite only having single-phase \textit{labelled} data, ideally, the network should be adapted for multi-phase operation on $\mathcal{D}_{u}$, meaning it should be able to consume whatever contrast phases are available and output a unified prediction that is stronger as more phases are available. 

To do this, we use \ac{HeMIS}-style feature fusion~\cite{havaei2016hemis}, which can predict masks given any arbitrary combination of contrast phases. To do this, a set of phase-specific layers produce a set of phase-specific activations, $\mathcal{A}$, whose cardinality depends on the number of inputs. The activations are then fused together using first- and second- order statistics, which are flexible enough to handle any number of inputs:
\begin{align}
    \mathbf{a}_{fuse} = \mathrm{concat}(\mu(\mathcal{A}), \mathrm{var}(\mathcal{A})) \mathrm{,}
\end{align}
where $\mathbf{a}_{fuse}$ denotes the fused feature, and the mean and variance are taken across the available phases. When only one phase is available, the variance features are set to $0$. To fuse intermediate predictions, an additional necessity for deeply-supervised networks, we simply take their mean. 

For choosing a fusion point, the co-training setup of \Sec~\ref{sec:co-train}, with its phase-specific layers, already offers a natural fusion location. We can then readily combine hetero-phase learning with co-training, re-defining a ``view'' to mean any possible combination of the four contrast phases. This has the added benefit of combinatorially increasing the number of co-training views. More formally, we use $\mathcal{X}^{*}=\mathcal{P}(\mathcal{X}) \setminus \{\varnothing\}$ to denote all possible contrast-phase combinations, where $\mathcal{P}(.)$ is the powerset operator. The corresponding predictions we denote as $\mathcal{\hat{Y}}^{*}$. When a data instance has all four phases, then the cardinality of  $\mathcal{X}^{*}$ and $\mathcal{\hat{Y}}^{*}$ is $15$, which is a drastic increase in views. With hetero-modality fusion in place, the consensus prediction and co-training loss of \eqref{eqn:M} and \eqref{eqn:co_instance_1}, respectively, can be supplanted by ones that use $\mathcal{\hat{Y}}^{*}$:
\begin{align}
M &= \dfrac{1}{|\mathcal{\hat{Y}}^{*}|}\sum_{\hat{Y}\in\mathcal{\hat{Y}}^{*}}\hat{Y} \mathrm{,} \label{eqn:M_2} \\
    \ell_{cons}(f(\mathcal{X})) &= \dfrac{1}{|\mathcal{\hat{Y}^{*}}|}\sum_{\hat{Y}\in \mathcal{\hat{Y}}^{*}}  \sum_{k \in \Omega}\mathit{KL}(\hat{Y}(k) \parallel M(k)) \mathrm{.} \label{eqn:co_instance_2}
\end{align}
When only single-phase combinations are used, \eqref{eqn:M_2} and \eqref{eqn:co_instance_2} reduce to standard co-training. To the best of our knowledge we are the first to combine co-training with hetero-modal learning. This combined workflow is graphically depicted in \Fig~\ref{fig:system}.

\subsection{Adversarial Domain Adaptation}

The co-heterogeneous training of \Sec~\ref{sec:co-hemis} is highly effective. Yet, it relies on accurate consensus predictions, which may struggle to handle significant appearance variations in $\mathcal{D}_{u}$ that are not represented in $\mathcal{D}_{\ell}$. Mask-based \ac{ADA}  offers an a complementary approach that trains a network to output masks that follow a \emph{prediction-based} distribution learned from labelled data~\cite{tsai2018learning}. Since liver shapes between $\mathcal{D}_{u}$ and $\mathcal{D}_{\ell}$ should follow similar distributions, this provides an effective learning strategy that is not as confounded by differences in appearance. Following Tsai \etal{}~\cite{tsai2018learning}, we can train a discriminator to classify whether a softmax output originates from a labelled- or unlabelled-dataset prediction. However, because we have a combinatorial number ($15$) of possible input phase combinations, \ie{} $\mathcal{\hat{X}}^{*}$, naively domain-adapting all corresponding predictions is prohibitively expensive. Fortunately, the formulations of \eqref{eqn:co_instance_1} and \eqref{eqn:co_instance_2} offer an effective and efficient solution. Namely, we can train the discriminator on the consensus prediction, $M$. This adapts the combinatorial number of possible predictions \textit{at the same computational cost as performing \ac{ADA} on only a single prediction}. 

More formally, let $d(.)$ be defined as an \ac{FCN} discriminator, then the discriminator loss can be expressed as
\begin{align}
    \mathcal{L}_{d} = \dfrac{1}{N_{\ell}}\sum_{\mathcal{D}_{\ell}}\ell_{ce}(d(\hat{Y}^{V}), \mathbf{1}) + \dfrac{1}{N_{u}}\sum_{\mathcal{D}_{u}}\ell_{ce}(d(M, \mathbf{0})) \mathrm{,} \label{eqn:D}
\end{align}
where $\ell_{ce}$ represents a pixel-wise cross-entropy loss. The opposing labels pushes the discriminator to differentiate semi-supervised consensus predictions from fully-supervised variants. Unlike natural image \ac{ADA}~\cite{tsai2018learning}, we do not wish to naively train the discriminator on all output classes, as it not reasonable to expect similar distributions of liver \textit{lesion} shapes across datasets. Instead we train the discriminator on the \textit{liver region}, \ie{} the union of healthy liver and lesion tissue predictions. Finally, when minimizing \eqref{eqn:D}, we only optimize the discriminator weights. The segmentation network can now be tasked with fooling the discriminator, through the addition of an adversarial loss:
\begin{align}
    \mathcal{L}_{adv} = \dfrac{1}{N_{u}}\sum_{\mathcal{D}_{u}}\ell_{ce}(d(M, \mathbf{1})) \mathrm{,} \label{eqn:l_adv}
\end{align}
where the ground-truth labels for $\ell_{ce}$ have been flipped from \eqref{eqn:D}. Note that here we use single-level \ac{ADA} as we found the multi-level variant~\cite{tsai2018learning} failed to offer significant enough improvements to offset the added complexity. When minimizing \eqref{eqn:l_adv}, or \eqref{eqn:total} for that matter, the discriminator weights are frozen. We empirically set $\lambda_{adv}$ to  $0.001$.

\subsection{Pseudo-Labelling}


By integrating co-heterogeneous training and \ac{ADA}, \ac{CHASe} can robustly segment challenging multi-phase unlabelled data. However some scenarios still present challenging edge cases, \eg{} lesions treated with \ac{TACE}. See the supplementary for some visualizations. To manage these cases, we use a simple, but effective, domain-specific pseudo-labelling. 

First, after convergence of \eqref{eqn:total}, we produce predictions on $\mathcal{D}_{u}$ using all available phases and extract any resulting 3D holes in the liver region (healthy tissue plus lesion) greater than $100$ voxels. Since there should never be 3D holes, these are mistakes. Under the assumption that healthy tissue in both datasets should equally represented, we treat these holes as missing ``lesion'' predictions. We can then create a pseudo-label, $Y_{h}$, that indicates lesion at the hole, \textit{with all others regions being ignored}. This produces a new ``holes'' dataset, $\mathcal{D}_{h}=\{\mathcal{X}, Y_{h}\}_{i=1}^{N_{h}}$, using image sets extracted from $\mathcal{D}_{u}$. We then finetune the model using \eqref{eqn:total}, but replace the segmentation loss of \eqref{eqn:seg} by
\begin{align}
    \mathcal{L}_{seg} =& \dfrac{1}{N_{\ell}}\sum_{V,Y\in\mathcal{D}_{\ell}} \ell_{seg}(f(V), Y) \nonumber \\
    &+ \dfrac{\lambda_{h}}{N_{h}}\sum_{\mathcal{X},Y_{h}\in\mathcal{D}_{h}}\sum_{X\in\mathcal{X}^{*}} \ell_{seg}(f(X), Y_{h}) \label{eqn:seg2}   \mathrm{,}
\end{align}
where we empirically set $\lambda_{h}$ to $0.01$ for all experiments. We found results were not sensitive to this value. While the hole-based pseudo-labels do not capture all errors, they only have to capture enough of missing appearances to guide \ac{CHASe}'s training to better handle recalcitrant edge cases.

\section{Results}

{\bf Datasets.} To execute \ac{CHASe}, we require datasets of single-phase labelled and multi-phase unlabelled studies, $\mathcal{D}_{u}$ and $\mathcal{D}_{\ell}$, respectively. The goal is to robustly segment patient studies from $\mathcal{D}_{u}$ while only having training mask labels from the less representative $\mathcal{D}_{\ell}$ dataset. 1) For $\mathcal{D}_{u}$, we collected $1147$ multi-phase dynamic \ac{CT} studies ($4577$ volumes in total) directly from the \ac{PACS} of \ac{CGMH}. The only selection criteria were patients with biopsied or resected liver lesions, with dynamic contrast \acp{CT} taken within one month before the procedure. Patients may have \ac{ICC}, \ac{HCC}, benign or metastasized tumors, along with co-occuring maladies, such as liver fibrosis, splenomegaly, or \ac{TACE}-treated tumors. Thus, $\mathcal{D}_{u}$ directly reflects the variability found in clinical scenarios. We used the DEEDS algorithm~\cite{Heinrich2013MRFBasedDR} to correct any misalignments. 2) For $\mathcal{D}_{\ell}$, we collected $235$ $V$-phase \ac{CT} studies collected from as many public sources as we could locate~\cite{bilic_liver_2019,4781564,gibson_eli_2018_1169361,chaos}. This is a superset of the LiTS training data~\cite{bilic_liver_2019}, and consists of a mixture of healthy and pathological livers, with only \ac{HCC} and metastasis represented.

{\bf Evaluation Protocols.} 1) To evaluate performance on $\mathcal{D}_{u}$, $47$ and $100$ studies were randomly selected for validation and testing, respectively, with $90$ test studies having all four phases. Given the extreme care required for lesion annotation, \eg{} the four readers used in the LiTS dataset~\cite{bilic_liver_2019}, only the liver region, \ie{} union of healthy liver and lesion tissue, of the $\mathcal{D}_{u}$ evaluation sets was annotated by a clinician. For each patient study, this was performed independently for each phase, with a final \emph{study-wise} mask generated via majority voting. 2) We also evaluate whether the \ac{CHASe} strategy of learning from unlabelled data can also improve performance on  $\mathcal{D}_{\ell}$. To do this, we split $\mathcal{D}_{\ell}$, with $70\%$/$10\%$/$20\%$ for training, validation, and testing, respectively, resulting in $47$ test volumes. To measure \ac{CHASe}'s impact on \emph{lesion} segmentation, we use the $\mathcal{D}_{\ell}$ test set.

{\bf Backbone Network.} We used a 2D segmentation backbone, an effective choice for many organs~\cite{harrison2017progressive,li2018h,Roth_2015,Roth_2018}, due to its simplicity and efficiency. We opt for the popular ResNet-50~\cite{he2016deep}-based DeepLabv2~\cite{Chen_2016} network with \ac{PHNN}-style deep supervision. We also tried VGG-16~\cite{simonyan2014very}, which also performed well and its results can be found in the supplementary. To create 3D masks we simply stack 2D predictions. {\bf \ac{CHASe} training.} We randomly sample multi-phase slices and, from them, randomly sample four out of the $15$ phase combinations from $\mathcal{X}^{*}$ to stochastically minimize \eqref{eqn:co_instance_2} and \eqref{eqn:l_adv}. For standard co-training baselines, we sample all available phases to minimize \eqref{eqn:co_instance_1}. {\bf Discriminator Network.} We use an \ac{ASPP} layer, employing dilation rates of {1,2,3,4} with a kernel size of 3 and a leaky ReLU with negative slope $0.2$ as our activation function. After a $1\times1$ convolution, a sigmoid layer classifies whether a pixel belongs to the labelled or unlabelled dataset. \emph{Specific details on data pre-processing, learning rates and schedules can be found in the supplementary material.}

\begin{figure}[t]
\center
      \includegraphics[width=.7\linewidth]{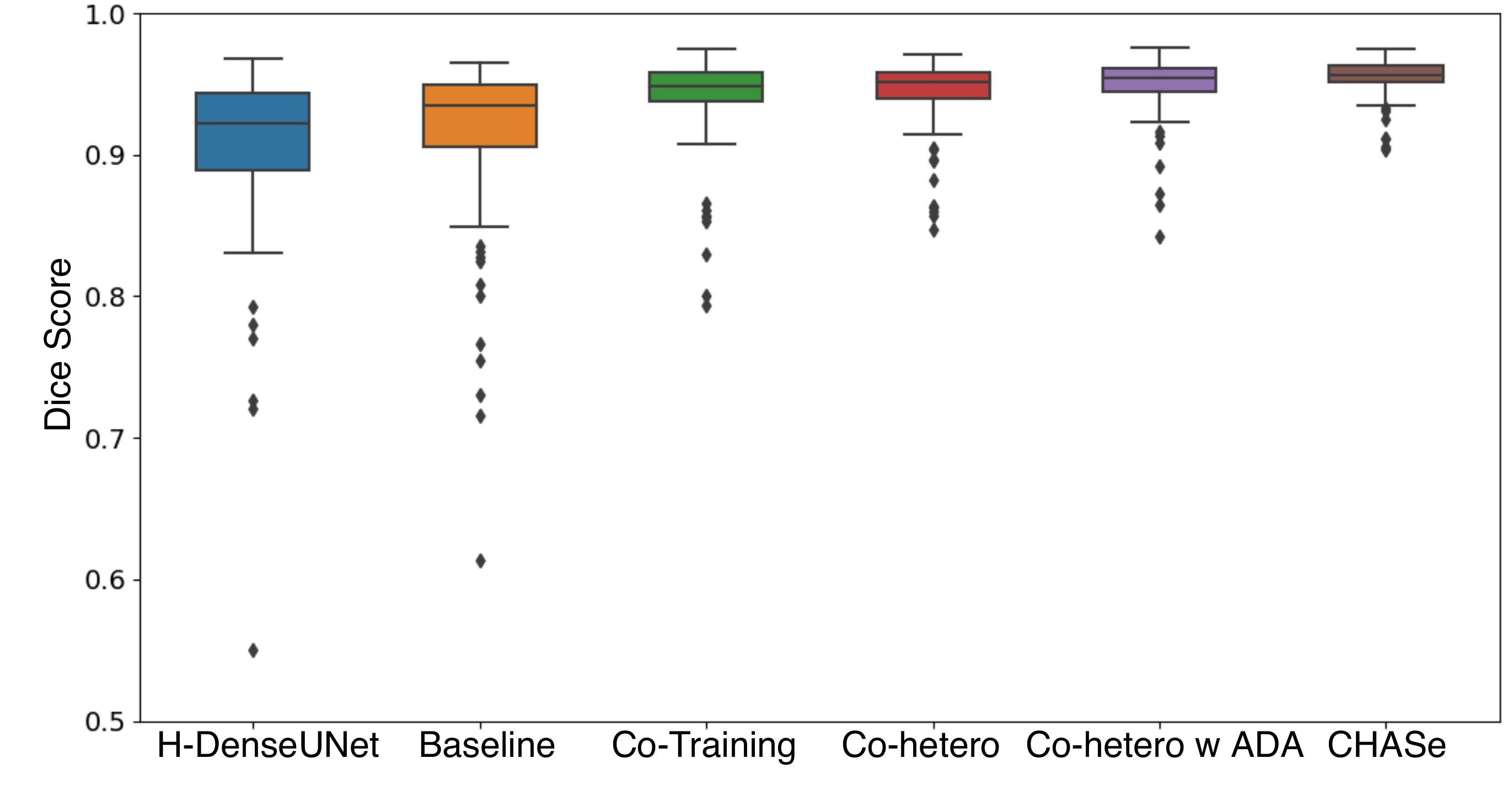}
    \caption{\small \textbf{Box and whisker plot.} Shown is the distribution of \acsp{DSC} of pathological liver segmentation on the \ac{CGMH} \ac{PACS}  when using \textit{all available} phases for inference.}
     \label{fig:box_plot}
\end{figure}

\subsection{Pathological Liver Segmentation}

We first measure the performance of \ac{CHASe} on segmenting pathological livers from the unlabeled \ac{CGMH} \ac{PACS} dataset, \ie{} $\mathcal{D}_{u}$. We use \ac{PHNN} trained only on $\mathcal{D}_{\ell}$ as a baseline, testing against different unlabeled learning baselines, \ie{} co-training~\cite{blum1998combining}, co-heterogeneous training, \ac{ADA}~\cite{tsai2018learning}, and hole-based pseudo-labelling. We measure the mean \ac{DSC} and \ac{ASSD}. For non hetero-modality variants, we use majority voting across each single-phase prediction to produce a multi-phase output. We also test against the publicly available hybrid H-DenseUNet model~\cite{li2018h}, one of the best published models. It uses a cascade of 2D and 3D networks.

As \Tab~\ref{tbl:cgmh_ablation} indicates, despite being only a single 2D network, our \ac{PHNN} baseline is strong, comparing similarly to the cascaded 2D/3D H-DenseUNet on our dataset\footnote{A caveat is that the public H-DenseUNet model was only trained on the LiTS subset of $\mathcal{D}_{\ell}$.}. 
\begin{table*}[t]
\caption{\textbf{Pathological Liver Segmentation.} Mean \ac{DSC} and \acs{ASSD} scores on the \ac{CGMH} \acs{PACS} dataset are tabulated across different contrast phase inputs. ``All'' means all available phases are used as input. Number of samples are  in parentheses.}
\small
\begin{center}

\begin{tabular}{|l|l|ll|ll|ll|ll|ll|}
\hline
\multicolumn{2}{|c|}{\multirow{2}{*}{Models}} & \multicolumn{2}{c|}{NC (96)}                             & \multicolumn{2}{c|}{A (98)}         & \multicolumn{2}{c|}{V (97)}         & \multicolumn{2}{c|}{D (98)}         & \multicolumn{2}{c|}{All (100)}       \\ \cline{3-12} 
\multicolumn{2}{|c|}{}                        & \multicolumn{1}{c|}{DSC} & \multicolumn{1}{c|}{ASSD} & \multicolumn{1}{c|}{DSC} & ASSD & \multicolumn{1}{c|}{DSC} & ASSD & \multicolumn{1}{l|}{DSC} & ASSD & \multicolumn{1}{l|}{DSC} & ASSD \\ \hline
\multicolumn{2}{|l|}{HDenseUNet~\cite{li2018h}}              &                          85.2&3.25                         &90.1&2.19
&90.7&2.61       &85.2&2.91     &                          89.9&2.59     \\
\multicolumn{2}{|l|}{Baseline~\cite{harrison2017progressive}}                &                          84.6&2.97                          &                         90.3 &1.23     &                         90.7 &1.18     &                          86.7&2.12    &                          91.4&1.21     \\
\multicolumn{2}{|l|}{Baseline w pseudo}          &                          89.4&1.97                          &                          90.5&1.34     &          90.9&1.29     &                          90.6&2.03     &                          91.9&1.27    \\
\multicolumn{2}{|l|}{Baseline w ADA~\cite{tsai2018learning}}            &                         90.9&1.34                          &                          91.9&1.13     &                          91.5&1.14     &                          90.9&1.65     &                         92.6&1.03    \\
\multicolumn{2}{|l|}{Co-training~\cite{qiao2018deep}}             &                     92.8&0.95                        &                     93.4&0.84    &                     93.4&0.83    &                          92.4&0.99     &                          94.0&0.92     \\
\multicolumn{2}{|l|}{Co-hetero}&                          93.4&0.81                          &                          93.7&0.77     &                         94.5&0.79     &                     93.6&0.86     &                         94.7 &0.89     \\
\multicolumn{2}{|l|}{Co-hetero w ADA}&                       93.8&0.81                         &                         93.9&0.79    &                     94.8&\textbf{0.66}    &                         93.9&0.81     &                         95.0&0.68     \\
\multicolumn{2}{|l|}{CHASe}&                        \textbf{94.0}&\textbf{0.79}                         &                          \textbf{94.2}&\textbf{0.74}     &                             \textbf{94.9}&\textbf{0.66}    &                                     \textbf{94.1}&\textbf{0.80}     &                        \textbf{95.4}&\textbf{0.63}     \\ \hline
\end{tabular}

 \end{center}{}
 \label{tbl:cgmh_ablation} 
\end{table*}
However, both H-DenseUNet and our \ac{PHNN} baseline  still struggle to perform well on the \ac{CGMH} dataset, particularly on non $V$-phases, indicating that training on public data alone is not sufficient. In contrast, through its principled semi-supervised approach, \ac{CHASe} is able to dramatically increase performance, producing boosts of $9.4\%$, $3.9\%$, $4.2\%$, $7.4\%$, and $4.0\%$ in mean \acp{DSC} for inputs of \ac{NC}, \ac{A}, \ac{V}, \ac{D}, and all phases, respectively. As can also be seen, all components contribute to these improvements, indicating the importance of each to the final result. Compared to established baselines of co-training and \ac{ADA}, \ac{CHASe} garners marked improvements. In addition, \ac{CHASe} performs more strongly as more phases are available, something the baseline models are not always able to do. Results across all $15$ possible combinations, found in our supplementary material, also demonstrate this trend.

More compelling results can be found in \Fig~\ref{fig:box_plot}'s box and whisker plots. As can be seen, each component is not only able to reduce variability, but more importantly significantly improves worst-case results. These same trends are seen across all possible phase combinations. Compared to improvements in mean \acp{DSC}, these worst-case reductions, with commensurate boosts in reliability, can often be more impactful for clinical applications. Unlike \ac{CHASe}, most prior work on pathological liver segmentation is fully-supervised. Wang \etal{} report $96.4\%$ \ac{DSC} on $26$ LiTS volumes and Yang \etal{}~\cite{yang2017automatic} report $95\%$ \ac{DSC} on $50$ test volumes with unclear healthy vs pathological status. We achieve comparable, or better, \acp{DSC} on $100$ pathological multi-phase test studies. As such, we articulate a versatile strategy to use and learn from the vast amounts of uncurated multi-phase clinical data housed within hospitals.

These quantitative results are supported by qualitative examples in \Fig~\ref{fig:results}. As the first two rows demonstrate, H-DenseUNet~\cite{li2018h} and our baseline can perform inconsistently across contrast phases, with both being confused by the splenomegaly (overly large spleen) of the patient. The \ac{CHASe} components are able to correct these issues. The third row in \Fig~\ref{fig:results} depicts an example of a \ac{TACE}-treated lesion, not seen in the public dataset and demonstrates how \ac{CHASe}'s components can progressively correct the under-segmentation. Finally, the last row depicts the \textit{worst-case} performance of \ac{CHASe}. Despite this unfavorable selection, \ac{CHASe} is still able to predict better masks than the alternatives. Of note, \ac{CHASe} is able to provide tangible improvements in consistency and reliability, robustly predicting even when presented with image features not seen in $\mathcal{D}_{\ell}$. More qualitative examples can be found in our supplementary material. 

\begin{figure*}[t]
\begin{center}
        \includegraphics[width=\linewidth]{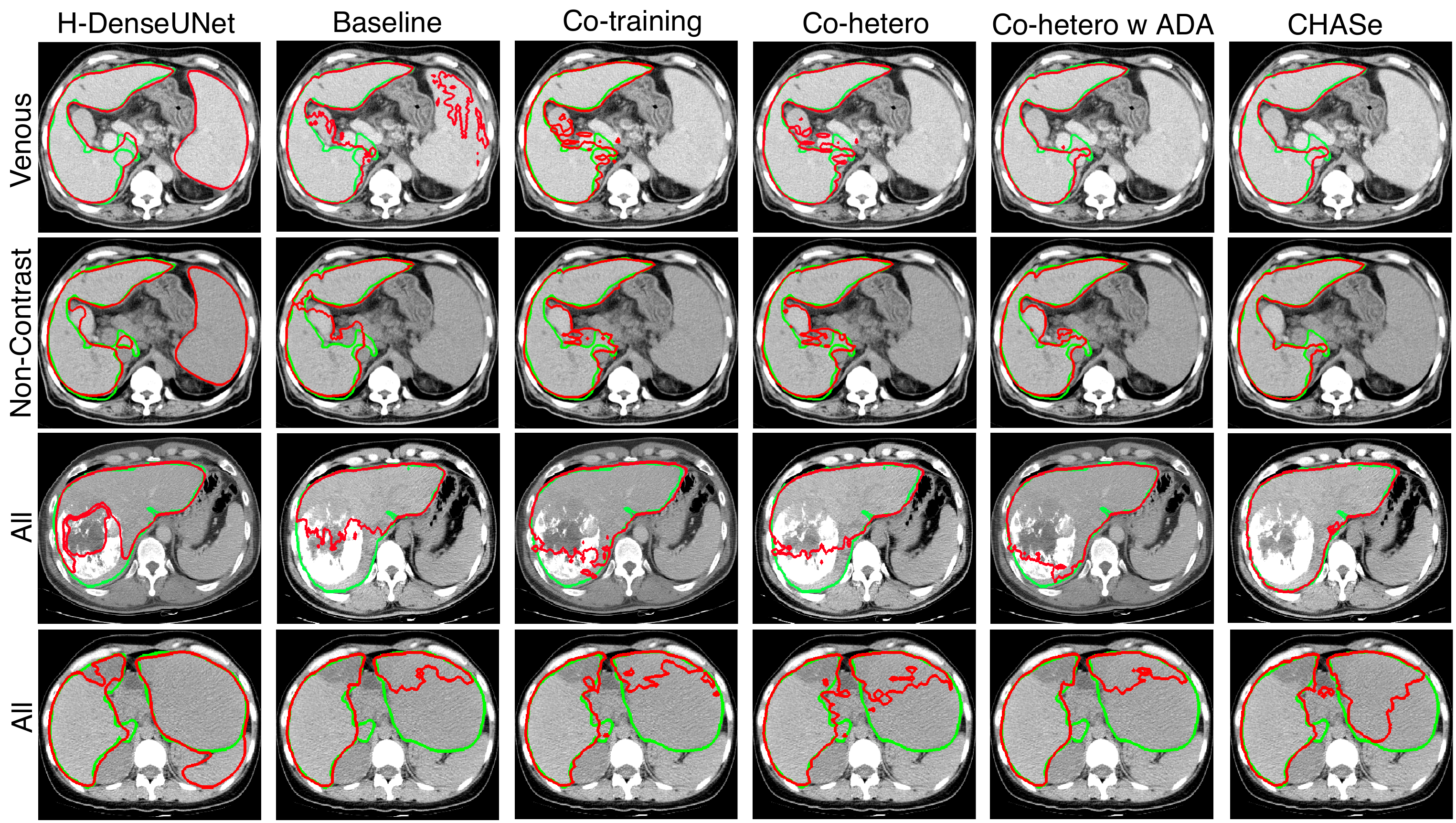}
\end{center}
\caption{\small \textbf{Qualitative results.}  The first two rows are captured from the \emph{same patient} across different contrast phases. The third and fourth row shows the performance when all available phases are included. {\bf Green} and {\bf red} curves depict the ground truth and segmentation predictions, respectively.}

\label{fig:results} 
\end{figure*}



\subsection{Liver and Lesion Segmentation}

We also investigate whether using \ac{CHASe} on unlabelled data can boost performance on the labelled data, \ie{} the $\mathcal{D}_{\ell}$ test set of $47$ single-phase \ac{V} volumes. Note, we include results from the public H-DenseUNet~\cite{li2018h} implementation, even though it was only trained on LiTS and included some of our test instances originating from LiTS in its training set.

\begin{table}[t]
\center
\caption{\textbf{Ablation study on public data.} Presented are test set \ac{DSC} scores with their standard deviation of healthy liver, lesion, and liver region.}
\small

\begin{tabular}{|l|l|l|l|l|l|l|l|}
\hline
\multicolumn{2}{|c|}{Model} & \multicolumn{2}{c|}{Liver} & \multicolumn{2}{c|}{Lesion} & \multicolumn{2}{l|}{Liver region} \\  \hline
\multicolumn{2}{|l|}{HDenseUNet~\cite{li2018h}}               & \multicolumn{2}{l|}{96.5 $\pm$ 2.0}      &\multicolumn{2}{l|}{51.7 $\pm$ 19.4}       & \multicolumn{2}{l|}{96.8 $\pm$ 1.8}  \\
\multicolumn{2}{|l|}{Baseline~\cite{harrison2017progressive}}               & \multicolumn{2}{l|}{96.3 $\pm$ 2.2}      &\multicolumn{2}{l|}{47.5 $\pm$ 24.1}       & \multicolumn{2}{l|}{96.6 $\pm$ 2.1}             \\
\multicolumn{2}{|l|}{Co-training}            & \multicolumn{2}{l|}{96.3 $\pm$ 1.8}      & \multicolumn{2}{l|}{51.9 $\pm$ 20.5}       & \multicolumn{2}{l|}{96.7 $\pm$ 1.7}             \\
\multicolumn{2}{|l|}{Co-hetero}               & \multicolumn{2}{l|}{96.4 $\pm$ 1.5}      & \multicolumn{2}{l|}{53.2 $\pm$ 19.1}       & \multicolumn{2}{l|}{96.7 $\pm$ 1.4}             \\
\multicolumn{2}{|l|}{Co-hetero w ADA}           & \multicolumn{2}{l|}{96.5 $\pm$ 1.5}      & \multicolumn{2}{l|}{\textbf{61.0 $\pm$ 17.2}}       & \multicolumn{2}{l|}{97.0 $\pm$ 1.3}             \\
\multicolumn{2}{|l|}{CHASe}                  & \multicolumn{2}{l|}{\textbf{96.8 $\pm$ 1.3}}      & \multicolumn{2}{l|}{60.3 $\pm$ 18.0}       & \multicolumn{2}{l|}{\textbf{97.1 $\pm$ 1.1}}             \\ \hline
\end{tabular}
\label{tbl:lesion} 
\end{table}

As \Tab~\ref{tbl:lesion} indicates, each \ac{CHASe} component progressively boosts performance, with lesion scores being the most dramatic. The one exception is that the holes-based pseudo-labelling produces a small decrease in mean lesion scores. Yet, box and whisker plots, included in our supplementary material, indicate that holes-based pseudo-labelling boosts \textit{median} values while reducing variability. Direct comparisons against other works, all typically using the LiTS challenge, are not possible, given the differences in evaluation data.  nnUNet~\cite{isensee2018nnu}, the winner of the Medical Decathlon, reported $61\%$ and $74\%$ \acp{DSC} for their own validation and challenge test set, respectively. However, $57\%$ of the patients in our test set are healthy, compared to the $3\%$ in LiTS. More healthy cases will tend to make it a harder lesion evaluation set, as any amount of false positives will produce \ac{DSC} scores of zero. For unhealthy cases, CHASe's lesion mean DSC is $61.9\%$ compared to $53.2\%$ for \ac{PHNN}. \ac{CHASe} allows a standard backbone, with no bells or whistles, to achieve dramatic boosts in lesion segmentation performance. As such, these results broaden the applicability of \ac{CHASe}, suggesting it can even improve the \textit{source}-domain performance of fully-supervised models.

\section{Conclusion}

We presented \ac{CHASe}, a powerful semi-supervised approach to organ segmentation. Clinical datasets often comprise multi-phase data and image features not represented in single-phase public datasets. Designed to manage this challenging domain shift, \ac{CHASe} can adapt publicly trained models to robustly segment multi-phase clinical datasets \textit{with no extra annotation}. To do this, we integrate co-training and hetero-modality into a co-heterogeneous training framework. Additionally, we propose a highly computationally efficient \ac{ADA} for multi-view setups and a principled holes-based pseudo-labeling. To validate our approach, we apply \ac{CHASe} to a highly challenging dataset of $1147$ multi-phase dynamic contrast \ac{CT} volumes of patients, all with liver lesions. Compared to strong fully-supervised baselines, \ac{CHASe} dramatically boosts mean performance ($>9\%$ in \ac{NC} \acp{DSC}), while also drastically improving worse-case scores. Future work should investigate 2.5D/3D backbones and apply this approach to other medical organs. Even so, these results indicate that \ac{CHASe} provides a powerful means to adapt publicly-trained models to challenging clinical datasets found ``in-the-wild''. 

\clearpage
%
%
\bibliographystyle{splncs04}
\bibliography{egbib}

\appendix
\section{Methodology Figures}

\begin{figure}
\center
      \includegraphics[width=75mm]{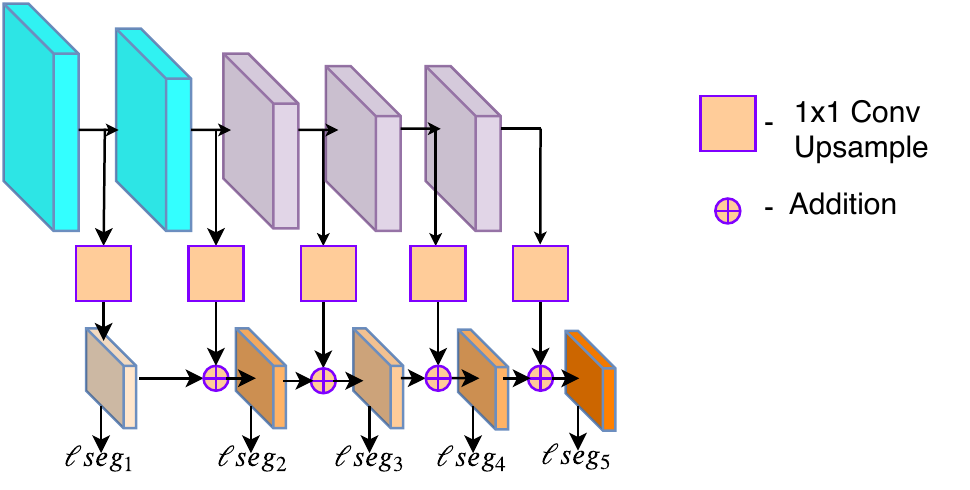}
    \caption{\small \textbf{PHNN architecture.} Here we use the \acs{V}-phase pathway coloring from \Fig~\ref{fig:system}. At each backbone stage, deeply supervised predictions and losses are calculated. Similar to residual-style connections~\cite{He2015DeepRL}, each stage's predictions are built off the prior one's using addition.}
     \label{fig:phnn}
\end{figure}

\begin{figure}
\center
      \includegraphics[width=\linewidth]{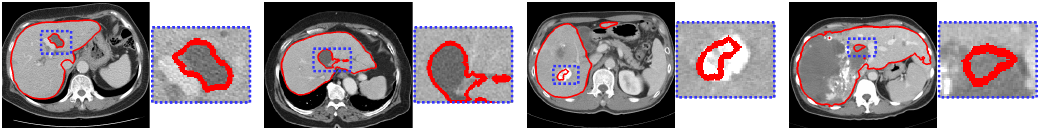}
    \caption{\small \textbf{Hole-Based Pseudo-Labelling.} 3D holes greater than $100$ voxels are extracted as lesion pseudo-masks missed by the prediction. Regions outside the hole are ignored. The third liver slice from left shows a \ac{TACE}-treated lesion, which is not seen in public datasets.}
     \label{fig:holes}
\end{figure}

Figures~\ref{fig:phnn} and \ref{fig:holes} depict the \ac{PHNN} architecture and our holes-based pseudo-labeling, respectively. 

\section{Implementation details}

\subsection{Network training}

We first initialize \ac{CHASe} with the weights trained on supervised venous phase data from the public datasets. To train this segmentation network, we use the Adam optimizer~\cite{DBLP:journals/corr/KingmaB14} with an initial learning rate of $3 \times 10^{-4}$ and values of $0.9$ and $0.99$ for the $\beta_{1}$ and $\beta_{2}$ hyperparamters, respectively. We reduce learning rateswhen the validation accuracy does not improve for $10$ epochs using a factor of $0.1$. 

To train \ac{CHASe}, we use the \ac{SGD} optimizer with an initial learning rate of $1 \times 10^{-5}$ and a momentum of $0.9$. We reduce the learning rate when the validation loss does not reduce for $10$ epochs using a factor of $0.1$. 

For training the discriminator, we use the Adam optimizer with an initial learning rate of $3 \times 10^{-4}$ and reduce the learning rate with a polynomial decay schedule with a power of $0.9$ as specified in~\cite{tsai2018learning}. 

We augment the dataset in both source and target domain by performing random rotation, random elastic deformation, gamma correction and random scaling.
\section{Additional Results}
\Tab~\ref{tbl:combi} shows the performance of different models on the test dataset using all $15$ possible combinations of phases during inference. For H-DenseUNet, Baseline, Co-training, which do not naturally accept multi-channel inputs, we perform majority voting across the appropriate single-phase predictions. 

\begin{table}
\caption{\textbf{Data distribution for $\mathcal{D}_{\ell}$.} Each dataset shows whether it contains only healthy liver or pathological liver and the number of volumes.}	
	\begin{center}		
		\begin{tabular}{ |p{2.1cm}|>{\centering}p{2.1cm}|>{\centering\arraybackslash}p{2.1cm}|>{\centering\arraybackslash}p{2.1cm}|}\hline
			{\bfseries Dataset ($\mathcal{D}_{\ell})$} & {\bfseries Total} & {\bfseries Healthy} & {\bfseries Pathological liver} \\ \hline
			LiTS & 130 &    & \cmark \\ \hline
			CHAOS  & 40 & \cmark & \\ \hline
			3D-IRCADb & 20 &   & \cmark\\ \hline
		    Gibson & 35 &   & \cmark \\ \hline
			Sliver07 & 20 & \cmark & \\ \hline
		\end{tabular}
	\end{center}
\label{tab:source_dataset_distribution}
\end{table}

\begin{table}[]
\caption{\textbf{Combination of views.} Mean \acp{DSC} are tabulated across different combinations of contrast phases used for input. The number of samples are indicated in parentheses. \cmark \hspace{0.05cm} signifies the presence of a phase and \xmark \hspace{0.05cm} represents the absence of a phase. }
\begin{tabular}{|llrl|llllll|}
\hline
\multicolumn{4}{|c|}{Multi-phases (90) } & \multicolumn{6}{c|}{Models}   \\ \hline
\multicolumn{1}{|c}{\rot{Non-contrast}} & \multicolumn{1}{c}{\rot{Arterial}} & \multicolumn{1}{c}{\rot{Venous}} & \multicolumn{1}{c|}{\rot{Delay}} & \multicolumn{1}{|c}{\rot{H-DenseUNet}} & \multicolumn{1}{c}{\rot{Baseline}} & \multicolumn{1}{c}{\rot{Co-training}} & \multicolumn{1}{c}{\rot{Co-hetero}} & \multicolumn{1}{c}{\rot{Cohetero w ADA}} & \multicolumn{1}{c|}{\rot{CHASe}} \\ \cline{1-10}
\xmark & \xmark  & \xmark & \cmark &85.7  &86.4  &92.9 &93.8  &94.0  &\textbf{94.3} \\ \cline{1-10}
\xmark & \xmark  & \cmark & \xmark & 90.9 &90.7  &93.7 &94.5  &94.9  &\textbf{95.0} \\ \cline{1-10}
\xmark & \xmark  & \cmark & \cmark & 90.5 &90.9  &93.8 &94.7  &\textbf{94.9}  &94.8 \\ \cline{1-10}
\xmark & \cmark  & \xmark & \xmark &90.8  &91.1  &93.6 &94.1  &94.3  &\textbf{94.6} \\ \cline{1-10}
\xmark & \cmark  & \xmark & \cmark &91.1  &91.3  &93.1 &94.6  &94.9  &\textbf{95.1} \\ \cline{1-10}
\xmark & \cmark  & \cmark & \xmark &91.9  &91.8  &92.9 &94.8  &94.8  &\textbf{95.0} \\ \cline{1-10}
\xmark & \cmark  & \cmark & \cmark &91.4  &91.6  &93.5 &95.0  &\textbf{95.2}  &\textbf{95.2} \\ \cline{1-10}
\cmark & \xmark  & \xmark & \xmark &85.6  &85.9  &92.4 &93.5  &93.8  &\textbf{94.0} \\ \cline{1-10}
\cmark & \xmark  & \xmark & \cmark &90.4  &90.7  &92.6 &93.8  &94.0  &\textbf{94.1} \\ \cline{1-10}
\cmark & \xmark  & \cmark & \xmark &91.1  &91.8  &93.4 &94.9  &95.1  &\textbf{95.2}\\ \cline{1-10}
\cmark & \xmark  & \cmark & \cmark &91.2  &92.0  &94.1 &94.8  &95.0  &\textbf{95.4}\\ \cline{1-10}
\cmark & \cmark  & \xmark & \xmark &90.9  &91.6  &93.7 &94.9  &94.8  &\textbf{95.0} \\ \cline{1-10}
\cmark & \cmark  & \xmark & \cmark &91.6  &91.4  &94.3 &95.0  &95.0  &\textbf{95.2} \\ \cline{1-10}
\cmark & \cmark  & \cmark & \xmark &91.5  &91.9  &94.4 &95.0  &95.1  &\textbf{95.3} \\ \cline{1-10}
\cmark & \cmark  & \cmark & \cmark &91.6  &92.1  &94.5 &95.1  &95.4  &\textbf{95.7} \\ \cline{1-10}
\end{tabular}
\label{tbl:combi}

\end{table}

\Tab~\ref{tbl:cgmh_ablation} provides the ablation study results when VGG16 is used as backbone. As can be seen, the results exhibit identical trends as when using a ResNet50-based DeepLabv2 backbone, except that absolute numbers are slightly worse. Nonetheless, even with an older backbone \ac{CHASe} is able to provide excellent results.

\begin{table*}[t]
\caption{\textbf{Pathological Liver Segmentation.} Mean \acs{DSC} and \acs{ASSD} results on the Anonymized \acs{PACS} dataset are tabulated across different contrast phase inputs. For ``All'', all available phases in the \ac{CT} study are used as input. Number of samples are indicated in parentheses. The segmentation model is trained with VGG16 backbone.}

\small
\begin{center}

\begin{tabular}{|l|l|ll|ll|ll|ll|ll|}
\hline
\multicolumn{2}{|c|}{\multirow{2}{*}{Models}} & \multicolumn{2}{c|}{NC (96)}                             & \multicolumn{2}{c|}{A (98)}         & \multicolumn{2}{c|}{V (97)}         & \multicolumn{2}{c|}{D (98)}         & \multicolumn{2}{c|}{All (100)}       \\ \cline{3-12} 
\multicolumn{2}{|c|}{}                        & \multicolumn{1}{c|}{DSC} & \multicolumn{1}{c|}{ASSD} & \multicolumn{1}{c|}{DSC} & ASSD & \multicolumn{1}{c|}{DSC} & ASSD & \multicolumn{1}{l|}{DSC} & ASSD & \multicolumn{1}{l|}{DSC} & ASSD \\ \hline
\multicolumn{2}{|l|}{HDenseUNet}              &                          85.2&3.25                        &90.1&2.19
&90.7&2.61       &85.2&2.91     &                          89.9&2.59     \\
\multicolumn{2}{|l|}{Baseline}                &                          85.1&2.81                          &                         90.1 &1.33     &                         90.2 &1.21     &                          86.9&2.03    &                          90.9&1.25     \\
\multicolumn{2}{|l|}{Baseline w pseudo}          &                          87.4&1.47                          &                          90.3&1.37     &          90.8&1.13     &                          91.1&1.12     &                          91.7&1.23    \\
\multicolumn{2}{|l|}{Baseline w ADA}            &                         88.3&1.38                          &                          91.2&1.08     &                          91.1&1.12     &                          92.1&0.99     &                         92.4&1.01    \\
\multicolumn{2}{|l|}{Co-training}             &                     91.8&1.03                        &                     92.5&1.01    &                     92.9&0.95    &                          92.5&1.02     &                         93.8&0.99     \\
\multicolumn{2}{|l|}{Co-hetero}&                          93.1&0.95                          &                          93.3&0.95     &                         94.0&0.80     &                     93.1&1.06     &                         94.6 &0.73     \\
\multicolumn{2}{|l|}{Co-hetero w ADA}&                       93.4&0.89                         &                         93.6&0.85    &                     \textbf{94.3}&0.74    &                         93.6&0.91     &                         94.7&0.73    \\
\multicolumn{2}{|l|}{CHASe}&                        \textbf{93.7}&\textbf{0.82}                        &                          \textbf{93.8}&\textbf{0.83}     &                             94.2&\textbf{0.73}    &                                     \textbf{93.8}&\textbf{0.87}     &                        \textbf{95.0}&\textbf{0.70}     \\ \hline
\end{tabular}

 \end{center}{}
 \label{tbl:cgmh_ablation}
 
\end{table*}

\begin{figure*}[t]
\begin{center}
        \includegraphics[width=.9\linewidth]{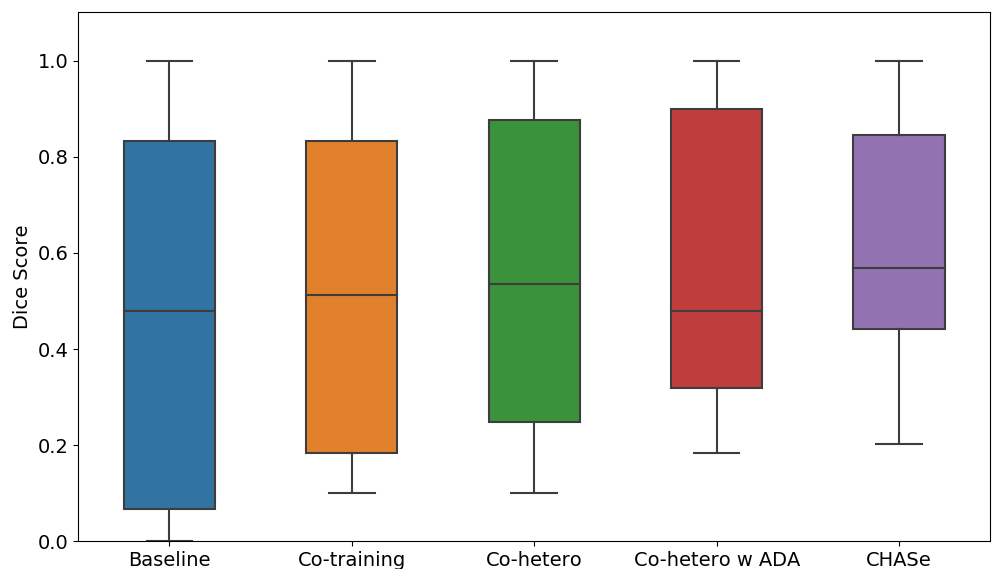}
\end{center}
\caption{Box-and-whisker plots of lesion scores on the public dataset. \acp{DSC} of $1.0$ and near $0.0$ are possible, as many studies had no lesions present. If the model did not predict any lesions, it yielded perfect \acp{DSC}. Conversely, predictions of any lesion when none are present penalize scores very heavily. }
\label{fig:system}
\end{figure*}

\Fig~\ref{fig:system} depicts a box-and-whisker plot of the lesion \ac{DSC} scores on the public dataset. As can be seen, all components of \ac{CHASe} contribute to higher performance.  Although the mean scores of \ac{CHASe} were lower when using the holes-based pseudo-labeling (see main text), the figure demonstrates that the median values are higher, with a tighter spread of quartile values. 

\begin{figure*}[t]
\begin{center}
        \includegraphics[width=\linewidth]{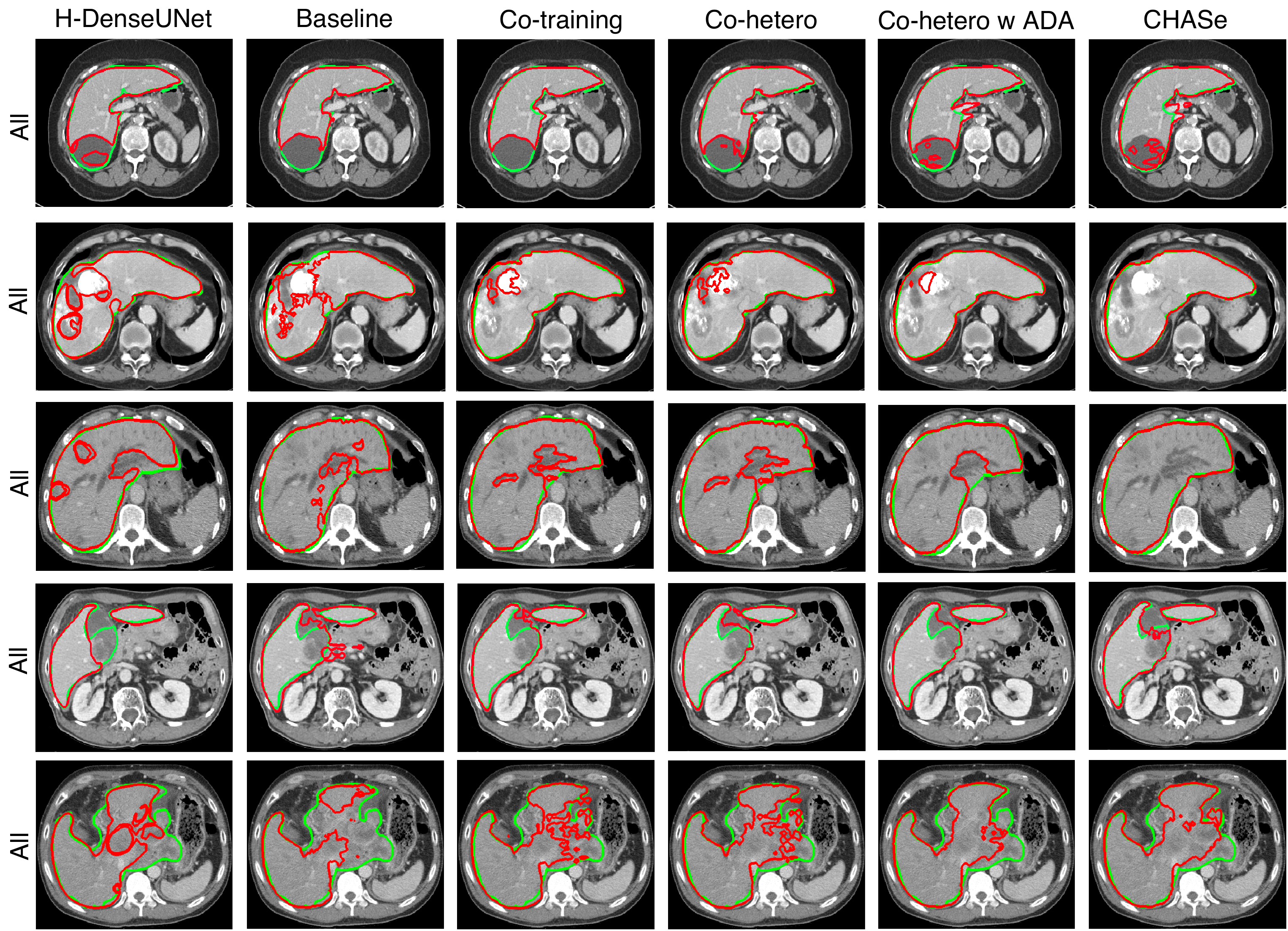}
\end{center}
\caption{\small \textbf{Qualitative results.}  {\bf Green} and {\bf red} curves depict the ground truth and segmentation predictions, respectively. All predictions executed with all phases used as input. The first and last rows depict failure cases, where the latter is an extremely challenging case with an extraordinarily large lesion occupying much of the liver space. \ac{CHASe} still manages to provide superior results compared to the alternatives. The second row demonstrates \ac{CHASe}'s ability to account for TACE-treated lesions, which are not present in public datasets. The fourth row depicts another highly challenging case, where the gallbladder is difficult to distinguish from a lesion. As can be seen, \ac{CHASe} is the only model able to successfully differentiate these two structures.}
\label{fig:results}

\end{figure*}

\Fig~\ref{fig:results} depicts additional qualitative results demonstrating the visual improvements provided by \ac{CHASe}.

\end{document}